\documentclass{aa}  
\usepackage{graphicx}
\usepackage{caption}
\usepackage{cuted}
\usepackage{subcaption}
\usepackage{xcolor}
\usepackage{soul}
\usepackage{dblfloatfix}
\usepackage{txfonts}
\usepackage{mathabx}
\usepackage[hidelinks]{hyperref}
\usepackage[normalem]{ulem}
\usepackage{orcidlink}
\usepackage{booktabs}
\usepackage{verbatim}
\usepackage{siunitx}
\usepackage{textcomp}
\usepackage{amsmath}
\usepackage{amsfonts}
\usepackage{amssymb}
\usepackage{float}

\newcounter{pta}

\DeclareMathSizes{12}{30}{16}{12}
\usepackage{amsmath,scalerel}

\makeatletter
\renewcommand*\aa@pageof{, page \thepage{} of \pageref*{LastPage}}
\makeatother

\usepackage{natbib}
\setcitestyle{authoryear,open={(},close={)}} 

\usepackage{titlesec}

\titlespacing\section{0pt}{14pt plus 2pt minus 2pt}{5pt plus 1pt minus 1pt}
\titlespacing\subsection{0pt}{12pt plus 4pt minus 2pt}{6pt plus 2pt minus 2pt}
\titlespacing\subsubsection{0pt}{12pt plus 4pt minus 2pt}{6pt plus 2pt minus 2pt}

\usepackage[switch]{lineno} 

\usepackage{upgreek}

\begin{document}

\title{RRAT-like behaviour of PSR~B0656$+$14 observed with I-LOFAR}

\author{
S.~C.~Susarla\orcidlink{0000-0003-4332-8201}\inst{\ref{uog}},
R.~Ferguson\orcidlink{0000-0002-9254-582X}\inst{\ref{uog}}
O.~A.~Johnson\orcidlink{0000-0002-5927-0481}\inst{\ref{tcd},\ref{ucb}},
L.~Vincetti\orcidlink{0009-0000-5589-0926}\inst{\ref{tcd}},
D.~J.~McKenna\orcidlink{0000-0001-7185-1310}\inst{\ref{astron}}, E.~F.~Keane\orcidlink{0000-0002-4553-655X}\inst{\ref{tcd}}, 
P.~J.~McCauley\orcidlink{0000-0003-4399-2233}\inst{\ref{tcd}},
A.~Golden\orcidlink{0000-0001-8208-4292}\inst{\ref{uog}}}
\institute{
Physics, School of Natural Sciences \& Center for Astronomy, College of Science and Engineering, University of Galway, University Road, Galway, H91 TK33, Ireland.\label{uog}\and
School of Physics, Trinity College Dublin, College Green, Dublin 2, D02 PN40, Ireland\label{tcd}\and
Radio Astronomy Laboratory, University of California, Berkeley, CA, USA\label{ucb}\and
ASTRON, The Netherlands Institute for Radio Astronomy, Oude Hoogeveensedijk 4, 7991 PD Dwingeloo, The Netherlands\label{astron}
}

   \date{Received XXX; accepted YYY}
\authorrunning{LOFAR}

 
  \abstract
{Single pulse studies offer vital insights into the emission physics of pulsars, particularly in the case of young, nearby sources where intrinsic variability is often pronounced. PSR~B0656+14, known for its sporadic and sometimes intense pulses, provides an excellent opportunity to investigate such behaviour at low radio frequencies.}
{This study aims to characterize the single pulse behaviour of PSR~B0656+14 using low-frequency observations at 110-190 MHz from the Irish LOFAR station. We focus on quantifying its pulse energy distribution, deriving precise dispersion measures (DMs) for both average and single pulses, analysing the temporal spacing (wait times) between subsequent pulses and studying the spectral indices for single pulses.}
{We employ standard pulsar timing techniques to derive the highest possible DM precision using integrated profiles. Single-pulse extraction is performed, and individual pulse DMs are estimated to probe pulse-to-pulse dispersion variability. We also perform a wait-time analysis to understand the statistical nature of pulse occurrence, and estimate the spectral index from frequency-resolved flux density measurements. The pulse energy distribution is modelled using a combination of log-normal and power-law components.}
{We report a DM of $14.053 \pm 0.005$~pc~cm$^{-3}$ for PSR~B0656+14 at LOFAR HBA frequencies. A total of 41 pulses were detected in a 5-hour observation, allowing a wait-time distribution analysis which is well-modelled by an exponential function, indicative of a Poisson process. Profile stability analysis indicates that a significant number of pulses (in excess of 47500) are required to reach a stable average profile, unusual compared to many other pulsars. The single-pulse spectral index varies significantly from pulse to pulse, with a mean value of $\alpha = -0.5$ and a standard deviation of $\Delta\alpha=1.3$. The pulse energy distribution shows a hybrid behaviour, consistent of a log-normal distribution and a power-law tail.}
{Our results confirm that PSR~B0656+14 exhibits highly variable, memory-less emission at low frequencies, with characteristics that resemble those seen in some rotating radio transients (RRATs). The need for an unusually large number of pulses to reach profile stability highlights the complex nature of its emission. As solar-wind studies demand high DM precision, this pulsar is insufficient for that purpose despite its proximity to the ecliptic plane—however SKA-Low's higher sensitivity could provide the required accuracy. While this study offers a detailed view of this pulsar's behaviour at LOFAR frequencies, further observations—both of this source and others—are essential. If such variability proves to be widespread among pulsars, population synthesis models and survey yield predictions would need to incorporate this currently overlooked feature to ensure accuracy.}

\keywords{}

   \maketitle



\section{Introduction}
Pulsars are rapidly rotating neutron stars that serve as highly stable cosmic clocks, primarily observable at radio frequencies \citep{LorimerandKramer2005}. These compact objects emit beams of electromagnetic radiation, and a pulse is detected when these intersect our line of sight. Pulsars are characterized by their remarkable periodicity, but can exhibit considerable variability in their emission \citep{brook_2019}, reflecting either intrinsic magnetospheric processes \citep{shaw_emission} or interactions with their local environment \citep{cordes1993,Zhong_2024}. Among the pulsars known to exhibit such complex emission behaviour is PSR~B0656+14 (or J0659+1414), first discovered by \citet{0656discovery} using the Molonglo Observatory Synthesis Telescope~\citep{molonglo}. It has a period of $385$~ms, a characteristic spin-down age $\tau\sim10^5$~yrs and is situated at a distance of $288\pm^{33}_{27}$~pc as determined through Very Long Baseline Interferometry~\citep{brisken_2003}. 
 
The emission behaviour of PSR~B0656+14 at higher frequencies closely resembles that of a rotating radio transient (RRAT)—a class of neutron stars that are more readily detected through their sporadic single pulses than through their phase-averaged periodic emission \citep{rrats2006,kk2011,zhang_2024}. RRATs are typically characterised by bright, yet intermittently detected, single pulses. The intriguing nature of PSR~B0656+14 was noted by \citet{walteverede2006b}, who pointed out that its relative proximity allows it to be detected through both its periodic emission and its RRAT-like bursts. It exhibits intense single pulses with high integrated energies and large peak flux densities, superimposed on a much weaker underlying periodic signal.

PSR~B0656+14 is of interest not only for its single pulse emission behaviour but also due to its proximity to the ecliptic plane, with an ecliptic latitude (ELAT) of $-8.4^\circ$. This location makes it a promising candidate for solar wind studies \citep{Tiburzi2021,susarla2024}, provided its dispersion measure can be determined with sufficient precision to trace variations induced by the Sun. In this paper, we present a detailed low-frequency (110–190 MHz) study of PSR~B0656+14, primarily focused on characterizing its single pulse emission. All observations were conducted using the Irish Low Frequency ARray (I-LOFAR) station~\citep{vanhaarlem2013} during dedicated standalone observing time. The structure of this paper is as follows: in Sect.~\ref{sec:obs}, we describe the telescope, observations, and data processing methods; in Sect.~\ref{sec:results}, we present and discuss our results; and finally, in Sect.~\ref{sec:disc}, we offer our conclusions.

\section{Observations}
\label{sec:obs}
\subsection{Data Acquisition}
Observations were conducted over 24 epochs in 2021 and 2022, accumulating a total observing time of $21.1$ hours. These were carried out using the high-band antenna (HBA) of I-LOFAR, operating in the $110-190$~MHz band. The data were initially analogue beam-formed, Nyquist-sampled with a 200-MHz clock, and coarse-channelised by the LOFAR backend hardware into 512 sub-bands. Subsequently, digital beam-forming was applied~\citep{vanhaarlem2013}. Of the $512$ resulting beam-formed complex voltage data streams, $488$ were recorded to disk at 8-bit precision~\citep{UDPDAVIDJOSS2023}. The data were then coherently dedispersed offline using the \textsc{cdmt} package \citep{cdmt2017}, and total-intensity Stokes $I$ \textsc{SigProc} filterbank files were generated~\citep{sigproc}.

The data were coherently dedispersed, using the conventional dispersion measure (DM) constant~\citep{kulkarni2020}, to a DM value of $13.94\;\mathrm{pc\,cm}^{-3}$, informed from the measurement of $13.94\pm0.09\;\mathrm{pc\,cm}^{-3}$ reported by \citet{pkj+13}, which was the reference value in the pulsar catalogue\footnote{\texttt{https://www.atnf.csiro.au/research/pulsar/psrcat/}}~\citep{psrcat} at the start of this project. This value has since been updated in the current version of the pulsar catalogue (\textsc{psrcat v2.5.1}) to $13.9\;\mathrm{pc\,cm}^{-3}$ (with no uncertainty specified), based on a measurement by \citet{whx+23}. Given LOFAR's low observing frequency, it is possible to achieve higher-precision DM measurements \citep{Donner202036msp}, which we discuss in the following sections.

\subsection{Data Processing}
\label{dataproc}
\noindent
Two different data processing methods were employed: 
\textit{(i) Folding Algorithm}: The final filterbanks have a central frequency of 149.902~MHz and were coherently dedispersed to this frequency. The data were then folded into 10-s sub-integrations using the \textsc{DSPSR} software suite \citep{vanstraaten2011}. The ephemeris used for folding was obtained from the ATNF pulsar catalogue \citep{psrcat}. Once the archives were created with these specifications, several post-processing steps were applied. During post-processing, each observation was cleaned of radio frequency interference (RFI) using a modified version of the \textsc{CoastGuard} software package~(see \citealt{lazaruscoast}; \href{https://www2.physik.uni-bielefeld.de/fileadmin/user_upload/radio_astronomy/Publications/Masterarbeit_LarsKuenkel.pdf}{Kuenkel 2017}\footnote{\url{https://github.com/larskuenkel/iterative_cleaner}}). Each observation was subsequently time-averaged and partially frequency-averaged into ten frequency sub-bands using the \textsc{PSRCHIVE} software suite~\citep{hotanpsrchive,straatenpsrchive}. The final bandwidth was restricted to $112-190$~MHz, as portions of the broader band were consistently affected by RFI and the system's high/low-pass filter edges. The epoch-wise DM time series was then generated following the procedure outlined in \citet{susarla2025}.

\begin{figure}[!h]
    \centering
    \includegraphics[width=\linewidth, height=1.1\linewidth]{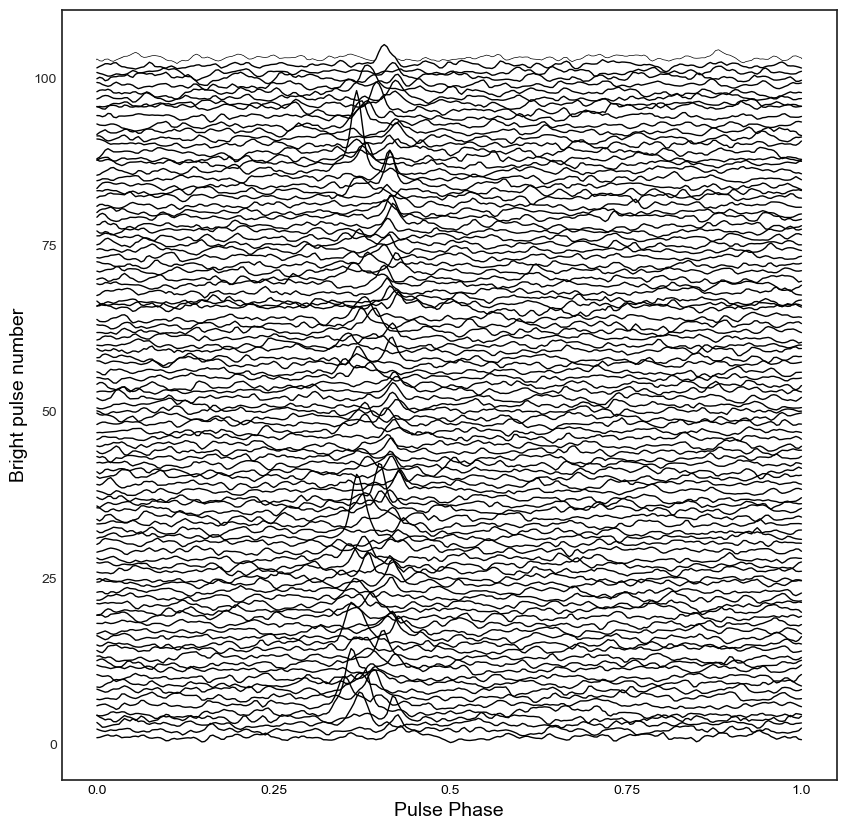}
    \caption{Waterfall plot of all the single pulses from 24 epochs taken in 2021-22.}
    \label{fig:jdplot}
\end{figure}

\textit{(ii) Single Pulse Analysis}: 
Using the processed filterbanks, the \texttt{rfifind} command in \textsc{PRESTO}~\citep{presto2011} was employed to identify and mask RFI-contaminated regions. The identified mask was then applied to the filterbanks before searching for single pulses using \textsc{TransientX}~\citep{transientx}. This approach was chosen for its efficiency in rapidly processing large datasets. Pulses were searched over a DM range of 0 to 25~$\mathrm{pc\,cm}^{-3}$ with a step size of 0.005~$\mathrm{pc\,cm}^{-3}$. To narrow down the search and increase sensitivity, an initial timing analysis was performed to determine a more precise reference DM, which was found to be approximately 14.05~$\mathrm{pc\,cm}^{-3}$ (see Sect.~\ref{sec: dm}). Based on this result, the search was refined to isolate pulses within the DM range of 13.9--14.2~$\mathrm{pc\,cm}^{-3}$. A detection threshold of $7\sigma$ above the noise floor was set to identify significant pulses, with each detection then also subject to visual inspection. This yielded a total of $103$ single pulses across the 24 epochs that were taken in 2021/22. These pulses are shown as a waterfall plot in Fig.~\ref{fig:jdplot}, with the per-epoch detection numbers summarized in Table~\ref{tab:spcount}.

\begin{table}[h]
    \centering
    \caption{Number of detected pulses ($N_{\rm p}$) for each MJD and their corresponding integration times ($T_{\rm int}$).}
    \begin{tabular}{ccc|ccc}
        \toprule
        \textbf{MJD} & {T$_{\rm int}$}(mins)&\textbf{$N_{\rm p}$} & \textbf{MJD} & {T$_{\rm int}$}(mins)& \textbf{$N_{\rm p}$} \\
        \hline
        59458 & 59 & 2  & 59556 & 59 & 5  \\
        59465 & 59 & 6  & 59563 & 59 & 7  \\
        59472 & 59 & 5  & 59570 & 59 & 4  \\
        59478 & 29 & 2  & 59641 & 59 & 4  \\
        59502 & 59 & 11 & 59647 & 59 & 7  \\
        59507 & 59 & 4  & 59653 & 59  & 6  \\
        59514 & 29 & 2  & 59660 & 44 & 6  \\
        59521 & 59 & 8  & 59667 & 29 & 3  \\
        59529 & 59 & 2  & 59674 & 34 & 1  \\
        59535 & 59 & 4  & 59681 & 29 & 1  \\
        59542 & 59 & 6  & 59688 & 29 & 1  \\
        59549 & 59 & 4  & 59695 & 29 & 2  \\
        \hline 
        \hline 
    \end{tabular}
    \label{tab:spcount}
\end{table}

\section{Results \& Discussions}
\label{sec:results}

This section presents a detailed analysis of the precise DMs obtained for the pulsar. Additionally, the statistical properties of the single-pulse analysis are examined, including the distribution of pulse energies. The wait-time distribution between successive pulses is investigated, including tests on profile stability, and the spectral indices of individual pulses are analysed.

\subsection{Dispersion measure}
\label{sec: dm}
\subsubsection{DMs of Folded epochs}
\label{sec: dmf}
\begin{figure}[h]
    \centering
    \includegraphics[width=0.95\linewidth]{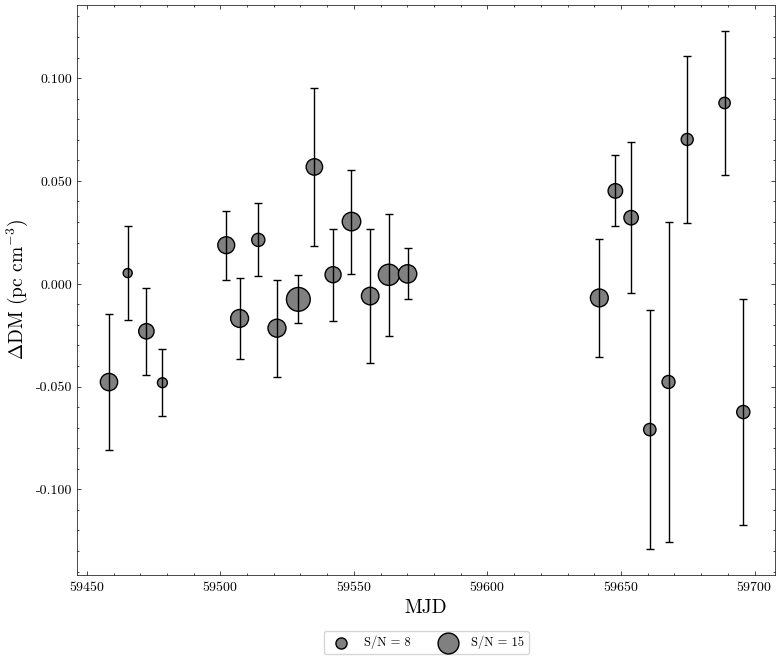}
    \caption{DM variation of PSR~B0656+14 as a function of MJD. Each DM value corresponds to a single observing epoch, derived using the epoch-wise method. A nominal DM of $14.053$~ pc cm$^{-3}$ has been subtracted from all the values. The circle sizes are representative of the S/N of the observation.}
    \label{fig:dmts}
\end{figure}

To obtain an accurate DM using I-LOFAR data, we employed the epoch-wise DM time-series method outlined in \citep{susarla2025}, following the approach detailed by \citet{iraci2024}. To create the template, all observations were combined to produce a high signal-to-noise ratio (S/N) profile. This was time-averaged, split into 10 frequency channels, smoothed individually, and finally summed and dedispersed. Using this template for the pulsar timing analysis, we determined a DM value of 14.053 $\pm$ 0.005 pc cm$^{-3}$. This measurement marks a significant improvement in precision, with uncertainties at least an order of magnitude smaller than those reported in \citet{pkj+13} and \citet{whx+23}. The individual DM measurements are shown in Fig.~\ref{fig:dmts}. The larger uncertainties in the final few observations are primarily due to shorter integration times as shown in Table. \ref{tab:spcount}. 

However, despite the precision of our DM determination, the uncertainties in individual measurements remain relatively large due to the pulsar’s intrinsically weak emission, in the context of the sensitivity of an international LOFAR station. For most observations, the folded S/N is less than $20$, which limits the effectiveness of timing analyses conducted over short durations. Consequently, this pulsar is not well-suited for timing studies based on one-hour observations with only a few bright pulses and weak emission otherwise, as it fails to achieve the stable pulse profile~\citep{walteverede2006b} formally necessary for precision pulsar timing~\citep{taylor1992}. A detailed analysis on profile stability is given in Sect.~\ref{sec: prof_stable}.

\subsubsection{DMs of single pulses}
To obtain the DM of single pulses, we used \textsc{dm\_phase}\footnote{\url{https://github.com/danielemichilli/DM_phase}} \citep{dm_phase2019}. \textsc{dm\_phase} determines the DM of a pulse by maximizing coherent power across the observed bandwidth, computing a coherence spectrum for a range of trial DM values by performing a 1D Fourier Transform along the frequency axis. Then, it normalizes the amplitude retaining only the phase information, and summing over all channels. The optimal DM is identified where the summed coherence spectrum is at a maximum. This ensures that dispersion-induced delays are corrected, and the pulse remains phase-aligned across frequencies.

\begin{figure}[!h]
    \centering
    \includegraphics[width=\linewidth]{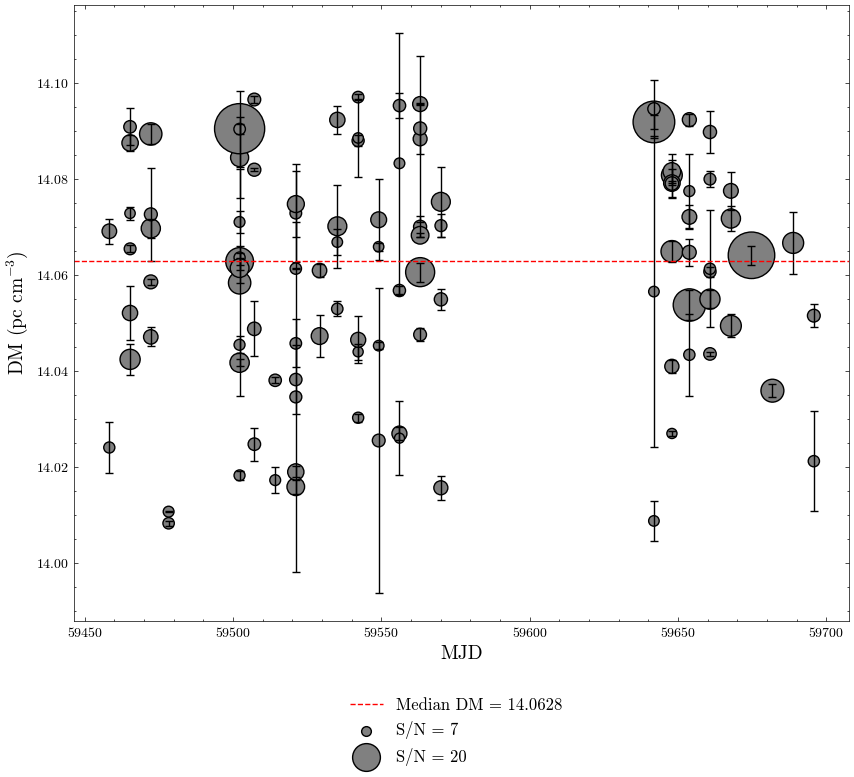}
    \caption{DM of the single pulses obtained using \textsc{DM\_phase} algorithm. The circle sizes are representative of the S/N of the pulse as shown in the legend. The red line is the median DM of all the single pulses.}
    \label{fig:spdms}
\end{figure}

For this work, we used the DM derived in Sect.~\ref{sec: dmf} as a reference (i.e., 14.053 pc cm$^{-3}$), and performed a DM search over the range 13.95-14.15 pc cm$^{-3}$. We adopted a DM step increment of $0.001$~pc cm$^{-3}$ which corresponds to the precision limit of our measurements given our measured S/N. This produced a grid of 200 trial DM values, for which \textsc{dm\_phase} estimated the DM for each of the single pulses. Fig.~\ref{fig:spdms} shows the DMs obtained for all the single pulses where the size of each circle represents the S/N value of the corresponding pulse. The pronounced variability in the DMs is likely driven by different spectral properties of individual pulses (see Sect.~\ref{sec: spindx}).

\begin{figure*}[!hb]
    \centering
    \includegraphics[width=1\linewidth]{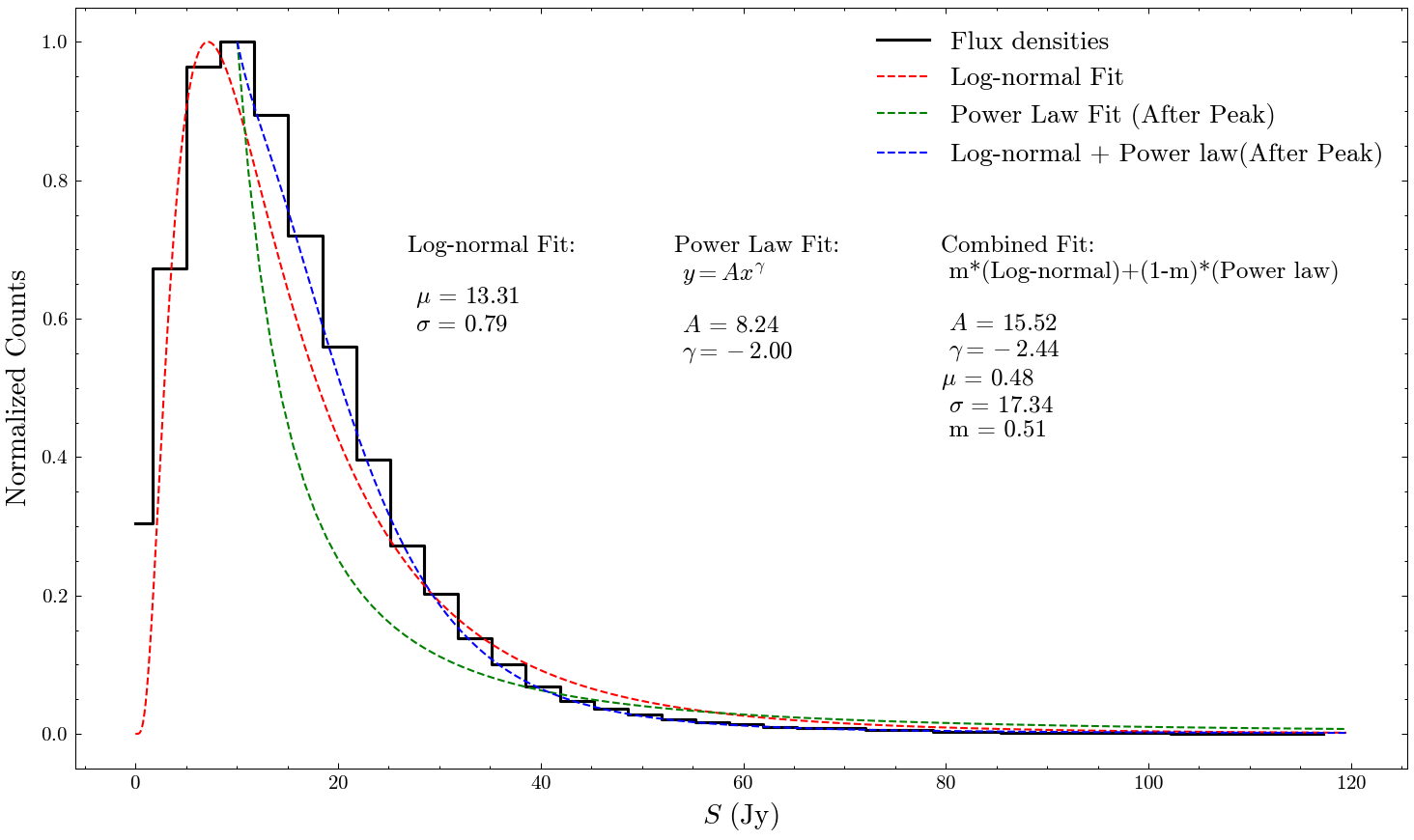}
    \caption{Noise incorporated flux density distribution of all the 103 pulses. The black line is the normalized histogram of the flux densities. We attempted to fit various models to this histogram. The red dashed line is the log-normal distribution and the green dashed line is the power-law distribution and the blue dashed line is the combined model of log-normal+ power law. The best fit model turned out to be a combination of 51\% log-normal and 49\% power law distribution.}
    \label{fig:fddist}
\end{figure*}

\begin{figure*}[!htp]
    \centering
    \includegraphics[width=\linewidth]{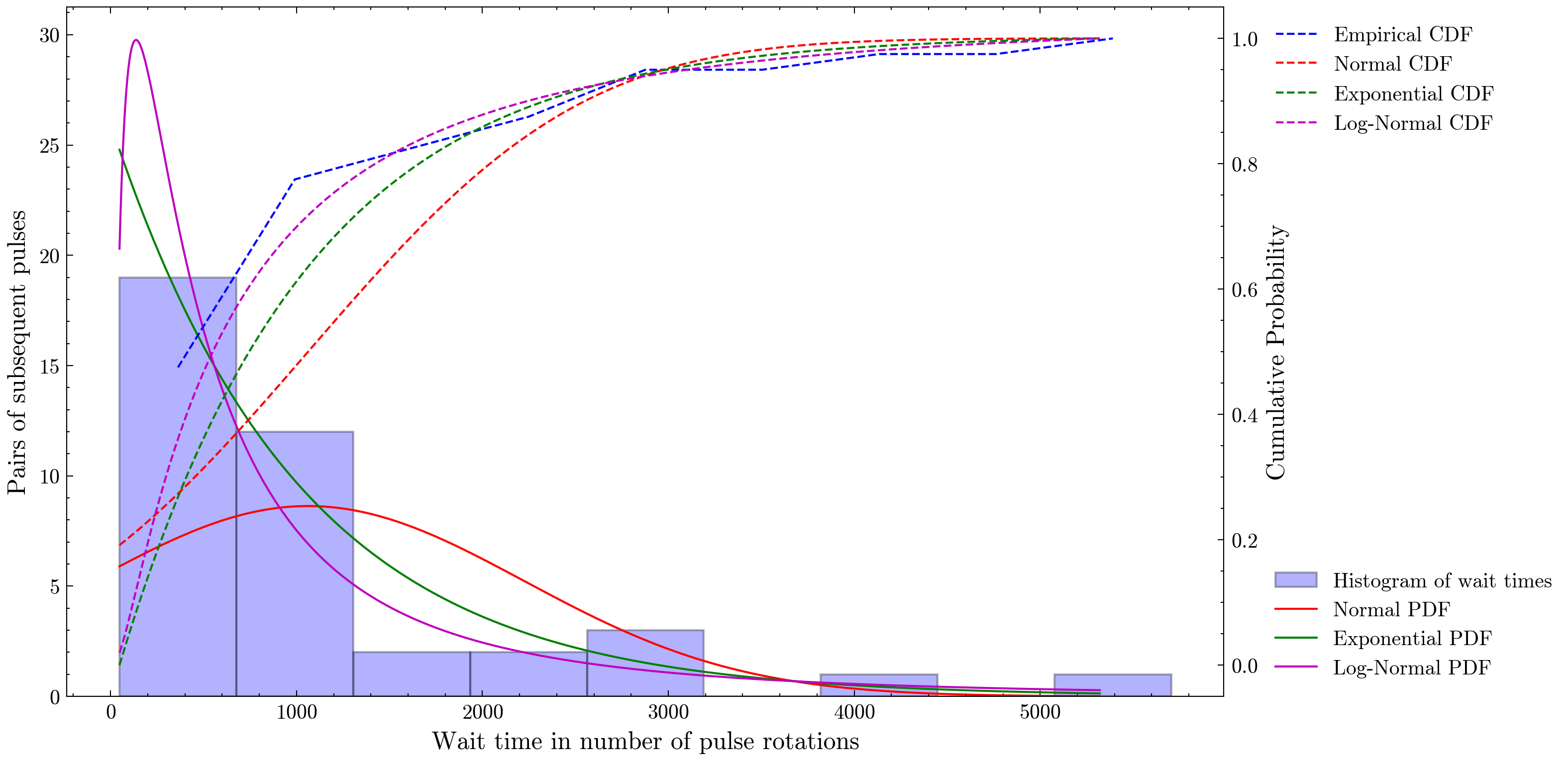}
    \caption{Wait time distribution of the single pulses observed on MJD 60747. The abscissa is the wait time between subsequent pulses expressed in terms of number of pulse periods. The ordinate is the number of pairs of subsequent pulses. The blue histogram is the distribution of the wait times. The solid lines represent the probability density functions (PDF) of various models and the dotted lines represent the cumulative distribution functions (CDF). The blue dotted line is the empirical CDF. It closely aligns with the exponential distribution implying a likely Poisson process. }
    \label{fig:wait_times}
\end{figure*}

\subsection{Flux density distribution}

The significant pulse-to-pulse variability of this pulsar makes it challenging to derive a meaningful longitude-resolved pulse amplitude spectrum. Additionally, the gain of a LOFAR station is inherently complex, varying with both zenith and azimuth angles (Vincetti et al., in prep.). As a result, the flux density scale fluctuates throughout our observations. To correct for these variations, we employ two methods, (a) we apply a Mueller matrix~\citep{mueller1943} correction factor, $M_{\rm II}$, from the radio interferometer measurement equation derived under the assumption of a beam model~\citep{hamaker2011,dreambeam}; and (b) estimating the sky temperature, $T_{\rm sky}$, by integrating over a sky model using a two-dimensional Gaussian convolution kernel that approximates the main lobe of the beam~\citep{david_rrats}. 
We utilize the \textsc{dreambeam}\footnote{\url{https://github.com/2baOrNot2ba/dreamBeam/tree/master}} software package to compute the Mueller matrix correction factor. We use the following equations to derive the flux densities for individual pulses and integrated profile,

\begin{align}
    S_{\rm pulse} &= \frac{2k_{\rm B}T_{\text{sys}}}{A_{\text{eff}}^{\text{max}} M_{\rm II} \sqrt{n_{\rm p} \Delta\nu\; w_{\text{pulse}}}} \times S/N_{\rm pulse}, \\
    S_{\rm mean} &= \frac{\beta}{n_{\text{chan}}} \sum_\nu \left( \frac{2 k_{\rm B} T_{\text{sys}} \times S/N_{\rm mean}}{A_{\text{eff}}^{\text{max}} M_{\rm II} \sqrt{n_{\rm p} \Delta\nu\; t_{\text{int}}}} \sqrt{\frac{W}{P - W}} \right).
\end{align}

\noindent where $T_{\rm sys}$ is the system temperature set at the convolved temperature of $373.12$~K at 150~MHz~\citep{lfss_dowell}, $A_{\text{eff}}^{\text{max}}$ is the maximum zenith effective area of the HBA at 150~MHz which we set at 2048 m$^2$~\citep{vanhaarlem2013}, $n_{\rm p}$ is the number of polarizations, $\Delta\nu$ is the channel bandwidth and $w_{\text{pulse}}$ is the pulse width. In Eq. 2, $t_{int}$ is the integration time of the observation. $W$ and $P$ are the pulse width and period respectively. Since most parameters in the above equations are frequency-dependent, we evaluate all quantities at a representative frequency of 150~MHz for simplicity. In our case, S/N and the $w_{\text{pulse}}$ for single pulses are obtained using the \textsc{TransientX} software. We note that the brightest individual pulse has $\sim$51~Jy, corresponding to a luminosity of $4.2$~Jy~kpc$^2$. This luminosity is on the same order of several other RRATs as detailed in \citet{0656rrat}. To incorporate receiver noise into our flux density measurements, we modeled the noise contribution as a Gaussian distribution with a standard deviation equal to 50\% of the flux density, following the approach of \citet{kondratiev_2016}. For each flux density value, we convolved this Gaussian noise and generated the resulting histogram shown in Fig.~\ref{fig:fddist}, which presents the noise-perturbed flux densities along with fitted distribution models. The number of bins are selected according to the \citet{fd81} rule which selects the optimum bin width based on the inter-quartile range of the data; we use this rule for all the histograms presented in this study. We explored three different fitting methods, namely log-normal, power-law and a combination of both. To evaluate the suitability of different emission energy distribution models, we compared single-component (log-normal, power-law) and combined (log-normal + power-law) fits with different admixture ratios using the \texttt{curve\_fit} function in \textsc{scipy} module. This performs a least-squares minimization for different free parameters and arrives at the best fit overall. Given the limited number of bins, we assessed the goodness-of-fit using the Poisson log-likelihood method. The log-likelihood scores indicate that the combined model ($\log \mathcal{L} = -54.71$) provides a significantly better fit than either the power-law ($\log \mathcal{L} = -131.33$) or the log-normal model ($\log \mathcal{L} = -1042.14$). To further quantify model performance, we applied the Akaike Information Criterion (AIC) and Bayesian Information Criterion (BIC), both of which favoured the combined model. This best-fit combined distribution consists of approximately 51\% log-normal and 49\% power-law contributions. The inferred power-law index for the combined model is $-2.44$, slightly steeper than the single power-law fit which yielded an index of $-2.00$. 

These results suggest a mixed-origin emission scenario, where the long tail in the flux density distribution is better explained by a hybrid population. For context, \citet{mckee2019} demonstrated that the giant pulse distribution of PSR~B1937+21 follows a broken power-law, while \citet{bsa2018rrat} found that RRATs in their study were best described by a log-normal distribution, with no improvement from adding a power-law component. \citet{waltevrede2006a} highlights the importance of multi-component fitting to address the simplistic fitting adopted in that work for higher frequency data of this pulsar. The substantial improvement seen in our combined model over the log-normal case—and moderate improvement over the pure power-law—indicates that the pulses in our sample may be more consistent with a giant pulse origin than with RRAT-like emission. Future observations with more sensitive low-frequency instruments, such as SKA-Low, hold the potential to detect a larger population of pulses, thereby providing clearer insight into the underlying emission mechanisms and distinguishing between RRAT-like and giant pulse-like behaviour.  

\subsection{Wait times analysis}
All observations conducted thus far have been limited to a maximum integration time of one hour, with the observation MJD 59502 yielding the highest number of detected pulses, totalling 11. Consequently, the number of available wait times between consecutive pulses is restricted to at best 10, which is insufficient for any meaningful statistical analysis. To address this limitation, we scheduled an extended observation of this pulsar for five hours on 13$^{th}$ March 2025 (MJD~60446). Using the same procedure outlined in Section \ref{dataproc}, we identified individual pulses, detecting a total of 41 pulses during this observation. 

\begin{table}[!h]
    \centering
    \caption{Kolmogorov-Smirnov test results for the Wait time distribution shown in Fig.~\ref{fig:wait_times}}
    \begin{tabular}{|c|c|c|} \hline 
         \textbf{Fit type}&  \textbf{Statistic}& \textbf{p-value}\\ \hline 
         \textbf{Normal}&  0.196& 0.08\\ \hline 
         \textbf{Exponential} &  0.134& 0.42\\ \hline 
         \textbf{log-normal}&  0.582& 2.2$\times$ 10$^{-13}$\\ \hline
    \end{tabular}
    \label{tab:kstest}
\end{table}

Fig.~\ref{fig:wait_times} shows the distribution of wait times of pairs of subsequent pulses in terms of number of pulse rotations. We attempted to model the distribution with a normal, exponential and log-normal distribution functions. In Fig.~\ref{fig:wait_times}, the solid lines are the probability density function and dotted lines are the cumulative distribution function. To check the goodness-of-fit, we also performed the Kolmogorov-Smirnov (K-S) test which checks whether the distribution follows the null hypotheses. The null hypothesis here is that the observed wait time distribution is drawn from a [normal / exponential / log-normal] distribution. The K-S test statistics are given in Table. \ref{tab:kstest}. From this analysis, we find that the exponential distribution provides the best fit to the observed wait times. The majority of pulses are emitted within a few hundred pulse periods of one another, consistent with a stochastic process. The wait time, on occasion, exceeds 5000 pulse rotations ($\sim$30 minutes). This supports the interpretation that the emission mechanism follows a Poisson process, suggesting that each pulse occurs independently of the others, without any underlying periodicity or memory of prior emission events.

\subsection{Profile stability}
\label{sec: prof_stable}
\begin{figure}[!h]
    \centering
    \includegraphics[width=\linewidth]{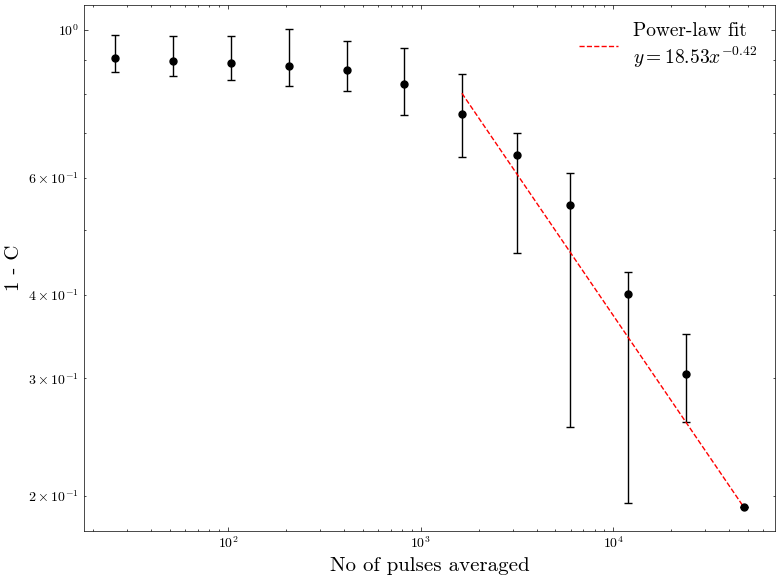}
    \caption{Cross-correlation factor plotted as (1–C) on the y-axis, where C represents the correlation between the template profile and sub-integrated averaged profiles. The x-axis shows the number of pulse rotations on a logarithmic scale. A lower (1–C) value indicates a higher similarity with the template. The red line denotes the best-fit power-law trend. The exponent for the power law fit is $-0.42\pm0.03$.}
    \label{fig:stability}
\end{figure}

To assess the stability of the average pulse profile at HBA frequencies, we utilized the $\sim$5-hour observation during which over 47500 pulse rotations were accumulated. We computed the correlation coefficient between the integrated pulse profile and a high-S/N template for varying sub-integration lengths, following the methods described by \citet{helfand1975} and \citet{rathnasree1995}. Fig.~\ref{fig:stability} presents the evolution of the correlation coefficient as a function of the number of averaged pulses. The results indicate a pronounced instability in the early phase of integration: the profile remains poorly correlated with the template up to at least 1000 pulse rotations. Beyond this point, a gradual improvement in correlation is observed, reaching a maximum value of only 0.80 after 47500 rotations. This slow convergence contrasts with the expected 1/N trend typically associated with stable pulsars \citep{liujitter}, N being the number of pulse rotations.

\begin{figure*}[htp]
    \centering
    \includegraphics[width=\linewidth, height=0.8\linewidth]{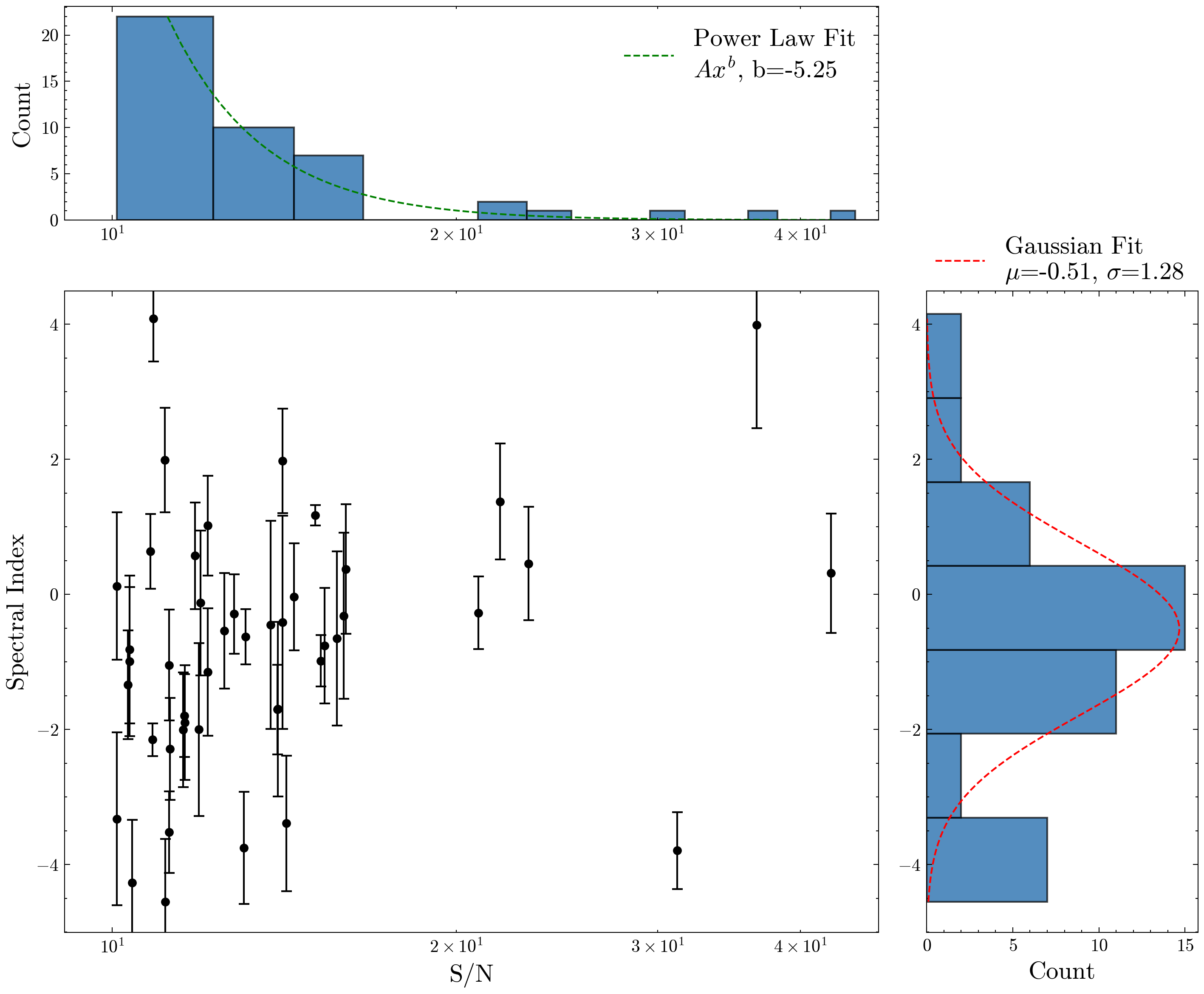}
    \caption{Figure showing the distribution of spectral indices and S/N for 45 pulses above 10$\sigma$ threshold. The top histogram is the distribution of S/N and the green dotted line is a power law fit with the legends highlighting the parameters. The vertical histogram on the left is the histogram of spectral indices. The red dotted line represents the best fitted Gaussian curve to the distribution. The mean spectral index is obtained as $\alpha=-0.5\pm1.3$.}
    \label{fig:spindxvS/N}
\end{figure*}

This behaviour contrasts sharply with well-studied pulsars like PSRs~B0329+54 and B1133+16, which reach comparable or higher correlation values after averaging just a few tens of pulses at 400 and 1400~MHz~\citep{helfand1975}. A similar trend was reported by \citet{walteverede2006b} that PSR~B0656+14 needed a large amount of pulses to achieve stability in the average pulse profile at 327 MHz, exceeding 25000 pulses corresponding to an observing time of 2.7 hours. They show that the spiky emission builds a narrow and peaked profile, whereas the weak emission produces a broad hump, which is largely responsible for the shoulders in the total emission profiles at both high and low frequencies.

\subsection{Single pulse spectral indices}
\label{sec: spindx}
In this section, we examine the spectral index variations in single pulses. To ensure robustness in our analysis, we restrict our sample to pulses that exceed an S/N of ten. This criterion is necessary because subdividing the frequency band into multiple subbands inherently reduces the S/N in each subband; for a flat spectrum and bandpass this would be by a factor of $\sqrt{n_{chan}}$ where $n_{chan}$ is the number of channels. Given the already faint nature of the pulsar at low frequencies, we limit our analysis to ten frequency channels per pulse. That gives us an approximate S/N for each channel as $(S/N)_{chan}=10/\sqrt{10}\sim3.2$. 

To determine the spectral indices, we employ the \texttt{pdmp} command from the \textsc{PSRCHIVE} software suite to obtain the S/N for each frequency channel. We then compute the corresponding flux densities using the radiometer equation (Eq. 1). The uncertainties on the flux densities are weighted according to the square of the S/N for each individual pulse. In reality, the flux densities could have an uncertainty scaled by up to 50\% \citep{kondratiev_2016}, for the LOFAR telescope. Assuming a power-law dependence of flux density on frequency, we express the spectral index $\alpha$ as:
\begin{equation*}
    S\propto \nu^\alpha\;,
\end{equation*}
where $S$ is the flux density at frequency $\nu$. Our weighted fit will weigh the subbands with the largest errorbars the least, ensuring that our final estimate remains statistically robust. The mean spectral index is then computed as the weighted mean of all single-pulse spectral indices. Figure \ref{fig:spindxvS/N} shows the distribution of spectral indices as a function of their S/N for all the single pulses. The spectral index shows significant variation, with a Gaussian fit yielding a mean of $\alpha = -0.5 \pm 1.3$.  This is broadly consistent with the spectral index of $\sim$ $\alpha = -0.5\pm0.2$ reported by \citet{lorimer1995} above 400 MHz for the integrated pulse profile. However, we note that the previous value by \citet{lorimer1995} has a small uncertainty as it was considered for the integrated pulse profiles rather than the single pulses. Our data exhibit a general trend where most pulses seem to be clustering around an index of zero suggesting that the flux is not changing much with an increase in frequency. We notice there is a significant pulse-to-pulse variability of spectral indices for this pulsar with a difference of $-8.5$ between the highest and lowest values. 

Several studies in the literature have explored the variability of single-pulse spectral indices across different pulsar populations. \citet{kramer2003fd} investigated two of the brightest pulsars, PSRs~B0329+54 and B1133+16, and found relatively modest variations in spectral indices, typically within a range of about $\sim$3. Interestingly, they noted that the individual components of giant pulses can exhibit slightly larger variations, up to $\sim$4. In contrast, \citet{karuppusamy2010} reported a much broader spread in the spectral indices of the Crab pulsar’s giant pulses, ranging from $-10$ to $+10$, reflecting the extreme variability of the Crab pulsar.

More recently, \citet{bsa2018rrat} presented a comprehensive study of single-pulse spectral indices for RRATs and Fast Radio Bursts (FRBs). Their results showed that RRATs typically exhibit spectral index variations on the order of $\sim$10, with FRBs often showing even wider spreads. Notably, one of the RRATs in their sample, J1930+1330, shows a spectral index distribution strikingly similar to that of PSR~B0656+14. \citet{zhang_2024} characterized J1930+1330 as a particularly intriguing source that shares emission characteristics with nulling pulsars, giant-pulse emitters, and even FRBs.

Given its wide spectral index spread, sporadic emission, and RRAT-like pulse morphology at low frequencies, PSR~B0656+14 likely exhibits RRAT-like behaviour in this regime. As \citet{0656rrat} suggest, if it were ~12 times more distant, it would probably be classified as an RRAT entirely.

\subsection{Solar wind}
\label{sw}
As previously noted, the ELAT of PSR~B0656+14 is $-8.45^{\circ}$ placing it relatively close to the ecliptic plane.
However, in practice, this pulsar is not ideally suited for such measurements. Based on the spherically symmetric model for solar wind-induced DM variations (Equation. 3 in \citealt{susarla2024}), the expected change in DM during its closest approach to the Sun is only $\sim$0.0008~pc~cm$^{-3}$. This level of variation is significantly smaller than the current observational precision, making it challenging to detect or characterize with existing data. 

During a 5-hour observation conducted on 13$^{\text{th}}$ March 2025, we obtained an integrated, folded DM precision of $\sim$0.009~pc~cm$^{-3}$, an order of magnitude larger than the required sensitivity to detect solar wind-induced DM variations.
Therefore, while PSR~B0656+14 lies in a geometrically favourable region, the combination of its relatively weak and intrinsically variable signal make it an impractical candidate for solar wind studies with current-generation telescopes. However, future instruments like SKA-Low, currently under construction in Western Australia, will have over an order of magnitude better relative sensitivity~\citep{keane2018}, which would enable these limitations to be overcome. 

\section{Conclusions}
\label{sec:disc}
In this work, we present a detailed single-pulse analysis of PSR~B0656+14 using observations obtained with the Irish LOFAR station in the frequency range 110–190 MHz. We report the most precise measurement to date of the pulsar’s DM, with a best-fit value of 14.053$\pm$0.005~pc~cm$^{-3}$. Despite this high precision, we find that it remains insufficient for several DM-based studies, including investigations of solar wind effects, as discussed in Sect.~\ref{sw}.

The pulsar exhibits sporadic emission, producing an average of approximately seven strong pulses per hour, while remaining weakly emitting otherwise at 150 MHz.
The flux density distribution exhibits a composite nature, best described by a mixture of log-normal (51\%) and power-law (49\%) components, with the power-law tail characterized by a spectral index of $-2.44$.
Additional insights were obtained from a dedicated 5-hour observation—significantly longer than our typical median pointing duration of 59 minutes—during which 41 single pulses were detected. The distribution of wait times between successive pulses follows an exponential trend, indicative of a Poisson process. This points toward a stochastic nature of the emission mechanism, with the brightest pulses occurring randomly rather than in any correlated or periodic pattern. This means that the pulsar is unlikely to be storing up energy for a fixed number of pulse rotations before emitting as bright bursts, or other processes involving `memory' of previous emission behaviour. Additionally, the profile stability analysis reveals that PSR~B0656+14 deviates from typical pulsar behaviour. The correlation coefficient of the average profile, with an analytic template derived from the entire dataset, remains low for the initial $\sim1000$ pulse rotations and increases only gradually with a power law exponent of $-0.42$, reaching a value of approximately 0.8 after $\sim$47,500 rotations, the duration of our longest observing epoch. This slow profile convergence implies that substantially longer integration times would be necessary to achieve sufficient profile stability for timing studies at low radio frequencies, i.e. such studies are impractical using this pulsar.

We also investigated the spectral indices of individual pulses, finding a broad distribution ranging from $-4$ to $+4$, with a weighted mean of $-$0.5$\pm$1.3. This highlights significant variability in the pulse spectra. While the spread of single pulse spectral indices is broader than that reported for PSRs~B0329+54 and B1133+16 by \citet{kramer2003fd}, it is narrower than the extreme variability observed in the giant pulses of the Crab pulsar \citep{karuppusamy2010}. The distribution bears a resemblance to those sources reported in \citet{bsa2018rrat}, where all cases exhibiting Gaussian-like profiles centred around negative spectral indices. The spread of spectral indices in particular is strikingly similar to RRAT~J1930+1330. This shows that PSR~B0656+14 behaves like an RRAT at these low frequencies. Future observations using next-generation telescopes, such as SKA-Low and the Five-hundred metre Aperture Spherical Telescope (FAST) when equipped with a VHF receiver~\citep{fast_dipole}, should offer the sensitivity required to further unravel the nature of this pulsar’s intriguing low-frequency radio emission.

\begin{acknowledgements}
SCS acknowledges the support of a University of Galway, College of Science and Engineering fellowship in supporting this work. OAJ acknowledges the support of Breakthrough Listen which is managed by the Breakthrough Prize Foundation. We thank Prof. Joris Verbiest for valuable discussions. The Rosse Observatory is operated by Trinity College Dublin. I-LOFAR infrastructure has benefited from funding from Science Foundation Ireland, a predecessor of Taighde \'{E}ireann --- Research Ireland. We thank the anonymous referee for their constructive comments which improved the quality of the paper. 
\end{acknowledgements}



\bibliographystyle{aa}
\bibliography{0656} 

\begin{thebibliography}{52}
\expandafter\ifx\csname natexlab\endcsname\relax\def\natexlab#1{#1}\fi

\bibitem[{{Bassa} {et~al.}(2017){Bassa}, {Pleunis}, \& {Hessels}}]{cdmt2017}
{Bassa}, C.~G., {Pleunis}, Z., \& {Hessels}, J.~W.~T. 2017, Astronomy and Computing, 18, 40

\bibitem[{{Brisken} {et~al.}(2003){Brisken}, {Thorsett}, {Golden}, \& {Goss}}]{brisken_2003}
{Brisken}, W.~F., {Thorsett}, S.~E., {Golden}, A., \& {Goss}, W.~M. 2003, \apjl, 593, L89

\bibitem[{{Brook} {et~al.}(2019){Brook}, {Karastergiou}, \& {Johnston}}]{brook_2019}
{Brook}, P.~R., {Karastergiou}, A., \& {Johnston}, S. 2019, \mnras, 488, 5702

\bibitem[{Carozzi(2020)}]{dreambeam}
Carozzi, T. 2020, {{dreamBeam}}

\bibitem[{{Cordes} {et~al.}(1993){Cordes}, {Romani}, \& {Lundgren}}]{cordes1993}
{Cordes}, J.~M., {Romani}, R.~W., \& {Lundgren}, S.~C. 1993, \nat, 362, 133

\bibitem[{{Donner} {et~al.}(2020){Donner}, {Verbiest}, {Tiburzi}, {Os{\l}owski}, {K{\"u}nsem{\"o}ller}, {Bak Nielsen}, {Grie{\ss}meier}, {Serylak}, {Kramer}, {Anderson}, {Wucknitz}, {Keane}, {Kondratiev}, {Sobey}, {McKee}, {Bilous}, {Breton}, {Br{\"u}ggen}, {Ciardi}, {Hoeft}, {van Leeuwen}, \& {Vocks}}]{Donner202036msp}
{Donner}, J.~Y., {et~al.}, {Tiburzi}, C., {Os{\l}owski}, S. 2020, \aap, 644, A153

\bibitem[{{Dowell} {et~al.}(2017){Dowell}, {Taylor}, {Schinzel}, {Kassim}, \& {Stovall}}]{lfss_dowell}
{Dowell}, J., {et~al.}, {Schinzel}, F.~K., {Kassim}, N.~E. 2017, \mnras, 469, 4537

\bibitem[{{Freedman} \& {Diaconis}(1981)}]{fd81}
{Freedman}, D. \& {Diaconis}, P. 1981, Zeitschrift fuer Wahrscheinlichkeitstheorie Werwandte Gebiete, 57, 453

\bibitem[{Hamaker(2011)}]{hamaker2011}
Hamaker, J. 2011, Mathematical-Physical Analysis of the Generic Dual-Dipole Antenna, Tech. rep., {Tech. rep., ASTRON}

\bibitem[{{Helfand} {et~al.}(1975){Helfand}, {Manchester}, \& {Taylor}}]{helfand1975}
{Helfand}, D.~J., {Manchester}, R.~N., \& {Taylor}, J.~H. 1975, \apj, 198, 661

\bibitem[{{Hotan} {et~al.}(2004){Hotan}, {van Straten}, \& {Manchester}}]{hotanpsrchive}
{Hotan}, A.~W., {van Straten}, W., \& {Manchester}, R.~N. 2004, \pasa, 21, 302

\bibitem[{{Iraci} {et~al.}(2024){Iraci}, {Chalumeau}, {Tiburzi}, {Verbiest}, {Possenti}, {Shaifullah}, {Susarla}, {Krishnakumar}, {Lam}, {Cromartie}, {Kerr}, \& {Grie{\ss}meier}}]{iraci2024}
{Iraci}, F., {et~al.}, {Tiburzi}, C., {Verbiest}, J.~P.~W. 2024, \aap, 692, A170

\bibitem[{{Karuppusamy} {et~al.}(2010){Karuppusamy}, {Stappers}, \& {van Straten}}]{karuppusamy2010}
{Karuppusamy}, R., {Stappers}, B.~W., \& {van Straten}, W. 2010, \aap, 515, A36

\bibitem[{{Keane}(2018)}]{keane2018}
{Keane}, E.~F. 2018, in IAU Symposium, Vol. 337, Pulsar Astrophysics the Next Fifty Years, ed. P.~{Weltevrede}, B.~B.~P. {Perera}, L.~L. {Preston}, \& S.~{Sanidas}, 158--164

\bibitem[{{Keane} \& {McLaughlin}(2011)}]{kk2011}
{Keane}, E.~F. \& {McLaughlin}, M.~A. 2011, Bulletin of the Astronomical Society of India, 39, 333

\bibitem[{{Kondratiev} {et~al.}(2016){Kondratiev}, {Verbiest}, {Hessels}, {Bilous}, {Stappers}, {Kramer}, {Keane}, {Noutsos}, {Os{\l}owski}, {Breton}, {Hassall}, {Alexov}, {Cooper}, {Falcke}, {Grie{\ss}meier}, {Karastergiou}, {Kuniyoshi}, {Pilia}, {Sobey}, {ter Veen}, {van Leeuwen}, {Weltevrede}, {Bell}, {Broderick}, {Corbel}, {Eisl{\"o}ffel}, {Markoff}, {Rowlinson}, {Swinbank}, {Wijers}, {Wijnands}, \& {Zarka}}]{kondratiev_2016}
{Kondratiev}, V.~I., {et~al.}, {Hessels}, J.~W.~T., {Bilous}, A.~V. 2016, \aap, 585, A128

\bibitem[{{Kramer} {et~al.}(2003){Kramer}, {Karastergiou}, {Gupta}, {Johnston}, {Bhat}, \& {Lyne}}]{kramer2003fd}
{Kramer}, M., {et~al.}, {Gupta}, Y., {Johnston}, S. 2003, \aap, 407, 655

\bibitem[{{Kulkarni}(2020)}]{kulkarni2020}
{Kulkarni}, S.~R. 2020, arXiv e-prints, arXiv:2007.02886

\bibitem[{{Lazarus} {et~al.}(2016){Lazarus}, {Karuppusamy}, {Graikou}, {Caballero}, {Champion}, {Lee}, {Verbiest}, \& {Kramer}}]{lazaruscoast}
{Lazarus}, P., {et~al.}, {Graikou}, E., {Caballero}, R.~N. 2016, \mnras, 458, 868

\bibitem[{{Liu} {et~al.}(2012){Liu}, {Keane}, {Lee}, {Kramer}, {Cordes}, \& {Purver}}]{liujitter}
{Liu}, K., {et~al.}, {Lee}, K.~J., {Kramer}, M. 2012, \mnras, 420, 361

\bibitem[{{Lorimer}(2011)}]{sigproc}
{Lorimer}, D.~R. 2011, {SIGPROC: Pulsar Signal Processing Programs}, Astrophysics Source Code Library, record ascl:1107.016

\bibitem[{{Lorimer} \& {Kramer}(2004)}]{LorimerandKramer2005}
{Lorimer}, D.~R. \& {Kramer}, M. 2004, {Handbook of Pulsar Astronomy} (Cambridge University Press)

\bibitem[{{Lorimer} {et~al.}(1995){Lorimer}, {Yates}, {Lyne}, \& {Gould}}]{lorimer1995}
{Lorimer}, D.~R., {Yates}, J.~A., {Lyne}, A.~G., \& {Gould}, D.~M. 1995, \mnras, 273, 411

\bibitem[{{Manchester} {et~al.}(2005){Manchester}, {Hobbs}, {Teoh}, \& {Hobbs}}]{psrcat}
{Manchester}, R.~N., {Hobbs}, G.~B., {Teoh}, A., \& {Hobbs}, M. 2005, \aj, 129, 1993

\bibitem[{{Manchester} {et~al.}(1978){Manchester}, {Lyne}, {Taylor}, {Durdin}, {Large}, \& {Little}}]{0656discovery}
{Manchester}, R.~N., {et~al.}, {Taylor}, J.~H., {Durdin}, J.~M. 1978, \mnras, 185, 409

\bibitem[{{McKee} {et~al.}(2019){McKee}, {Stappers}, {Bassa}, {Chen}, {Cognard}, {Gaikwad}, {Janssen}, {Karuppusamy}, {Kramer}, {Lee}, {Liu}, {Perrodin}, {Sanidas}, {Smits}, {Wang}, \& {Zhu}}]{mckee2019}
{McKee}, J.~W., {et~al.}, {Bassa}, C.~G., {Chen}, S. 2019, \mnras, 483, 4784

\bibitem[{{McKenna} {et~al.}(2023){McKenna}, {Keane}, {Gallagher}, \& {McCauley}}]{UDPDAVIDJOSS2023}
{McKenna}, D.~J., {Keane}, E.~F., {Gallagher}, P.~T., \& {McCauley}, J. 2023, arXiv e-prints, arXiv:2309.03228

\bibitem[{{McKenna} {et~al.}(2024){McKenna}, {Keane}, {Gallagher}, \& {McCauley}}]{david_rrats}
{McKenna}, D.~J., {Keane}, E.~F., {Gallagher}, P.~T., \& {McCauley}, J. 2024, \mnras, 527, 4397

\bibitem[{{McLaughlin} {et~al.}(2006){McLaughlin}, {Lyne}, {Lorimer}, {Kramer}, {Faulkner}, {Manchester}, {Cordes}, {Camilo}, {Possenti}, {Stairs}, {Hobbs}, {D'Amico}, {Burgay}, \& {O'Brien}}]{rrats2006}
{McLaughlin}, M.~A., {et~al.}, {Lorimer}, D.~R., {Kramer}, M. 2006, \nat, 439, 817

\bibitem[{{Men} \& {Barr}(2024)}]{transientx}
{Men}, Y. \& {Barr}, E. 2024, \aap, 683, A183

\bibitem[{{Mills}(1981)}]{molonglo}
{Mills}, B.~Y. 1981, \pasa, 4, 156

\bibitem[{Mueller(1943)}]{mueller1943}
Mueller, H. 1943, The polarization optics of the photoelectric shutter, Tech. rep., {United States: Office of Scientific Research and Development, National Defense Research Committee, Division 16---Optics and Camouflage, 11-15}

\bibitem[{{Petroff} {et~al.}(2013){Petroff}, {Keith}, {Johnston}, {van Straten}, \& {Shannon}}]{pkj+13}
{Petroff}, E., {et~al.}, {Johnston}, S., {van Straten}, W. 2013, \mnras, 435, 1610

\bibitem[{{Ransom}(2011)}]{presto2011}
{Ransom}, S. 2011, {PRESTO: PulsaR Exploration and Search TOolkit}, Astrophysics Source Code Library, record ascl:1107.017

\bibitem[{{Rathnasree} \& {Rankin}(1995)}]{rathnasree1995}
{Rathnasree}, N. \& {Rankin}, J.~M. 1995, \apj, 452, 814

\bibitem[{{Seymour} {et~al.}(2019){Seymour}, {Michilli}, \& {Pleunis}}]{dm_phase2019}
{Seymour}, A., {Michilli}, D., \& {Pleunis}, Z. 2019, {DM\_phase: Algorithm for correcting dispersion of radio signals}, Astrophysics Source Code Library, record ascl:1910.004

\bibitem[{{Shapiro-Albert} {et~al.}(2018){Shapiro-Albert}, {McLaughlin}, \& {Keane}}]{bsa2018rrat}
{Shapiro-Albert}, B.~J., {McLaughlin}, M.~A., \& {Keane}, E.~F. 2018, \apj, 866, 152

\bibitem[{{Shaw} {et~al.}(2022){Shaw}, {Stappers}, {Weltevrede}, {Brook}, {Karastergiou}, {Jordan}, {Keith}, {Kramer}, \& {Lyne}}]{shaw_emission}
{Shaw}, B., {et~al.}, {Weltevrede}, P., {Brook}, P.~R. 2022, \mnras, 513, 5861

\bibitem[{{Susarla} {et~al.}(2024){Susarla}, {Chalumeau}, {Tiburzi}, {Keane}, {Verbiest}, {Hazboun}, {Krishnakumar}, {Iraci}, {Shaifullah}, {Golden}, {Bak Nielsen}, {Donner}, {Grie{\ss}meier}, {Keith}, {Os{\l}owski}, {Porayko}, {Serylak}, {Anderson}, {Br{\"u}ggen}, {Ciardi}, {Dettmar}, {Hoeft}, {K{\"u}nsem{\"o}ller}, {Schwarz}, \& {Vocks}}]{susarla2024}
{Susarla}, S.~C., {et~al.}, {Tiburzi}, C., {Keane}, E.~F. 2024, \aap, 692, A18

\bibitem[{Susarla {et~al.}(2025)Susarla, Johnson, McKenna, Keane, McCauley, Verbiest, Tiburzi, \& Golden}]{susarla2025}
Susarla, S.~C., {et~al.}, McKenna, D.~J., Keane, E.~F. 2025 [\eprint[arXiv]{2505.09549}]

\bibitem[{{Taylor}(1992)}]{taylor1992}
{Taylor}, J.~H. 1992, Philosophical Transactions of the Royal Society of London Series A, 341, 117

\bibitem[{{Tiburzi} {et~al.}(2021){Tiburzi}, {Shaifullah}, {Bassa}, {Zucca}, {Verbiest}, {Porayko}, {van der Wateren}, {Fallows}, {Main}, {Janssen}, {Anderson}, {Bak Nielsen}, {Donner}, {Keane}, {K{\"u}nsem{\"o}ller}, {Os{\l}owski}, {Grie{\ss}meier}, {Serylak}, {Br{\"u}ggen}, {Ciardi}, {Dettmar}, {Hoeft}, {Kramer}, {Mann}, \& {Vocks}}]{Tiburzi2021}
{Tiburzi}, C., {et~al.}, {Bassa}, C.~G., {Zucca}, P. 2021, \aap, 647, A84

\bibitem[{{van Haarlem} {et~al.}(2013){van Haarlem}, {Wise}, {Gunst}, {Heald}, {McKean}, {Hessels}, {de Bruyn}, {Nijboer}, {Swinbank}, {Fallows}, {Brentjens}, {Nelles}, {Beck}, {Falcke}, {Fender}, {H{\"o}randel}, {Koopmans}, {Mann}, {Miley}, {R{\"o}ttgering}, {Stappers}, {Wijers}, {Zaroubi}, {van den Akker}, {Alexov}, {Anderson}, {Anderson}, {van Ardenne}, {Arts}, {Asgekar}, {Avruch}, {Batejat}, {B{\"a}hren}, {Bell}, {Bell}, {van Bemmel}, {Bennema}, {Bentum}, {Bernardi}, {Best}, {B{\^\i}rzan}, {Bonafede}, {Boonstra}, {Braun}, {Bregman}, {Breitling}, {van de Brink}, {Broderick}, {Broekema}, {Brouw}, {Br{\"u}ggen}, {Butcher}, {van Cappellen}, {Ciardi}, {Coenen}, {Conway}, {Coolen}, {Corstanje}, {Damstra}, {Davies}, {Deller}, {Dettmar}, {van Diepen}, {Dijkstra}, {Donker}, {Doorduin}, {Dromer}, {Drost}, {van Duin}, {Eisl{\"o}ffel}, {van Enst}, {Ferrari}, {Frieswijk}, {Gankema}, {Garrett}, {de Gasperin}, {Gerbers}, {de Geus}, {Grie{\ss}meier}, {Grit}, {Gruppen}, {Hamaker}, {Hassall}, {Hoeft}, {Holties},
  {Horneffer}, {van der Horst}, {van Houwelingen}, {Huijgen}, {Iacobelli}, {Intema}, {Jackson}, {Jelic}, {de Jong}, {Juette}, {Kant}, {Karastergiou}, {Koers}, {Kollen}, {Kondratiev}, {Kooistra}, {Koopman}, {Koster}, {Kuniyoshi}, {Kramer}, {Kuper}, {Lambropoulos}, {Law}, {van Leeuwen}, {Lemaitre}, {Loose}, {Maat}, {Macario}, {Markoff}, {Masters}, {McFadden}, {McKay-Bukowski}, {Meijering}, {Meulman}, {Mevius}, {Middelberg}, {Millenaar}, {Miller-Jones}, {Mohan}, {Mol}, {Morawietz}, {Morganti}, {Mulcahy}, {Mulder}, {Munk}, {Nieuwenhuis}, {van Nieuwpoort}, {Noordam}, {Norden}, {Noutsos}, {Offringa}, {Olofsson}, {Omar}, {Orr{\'u}}, {Overeem}, {Paas}, {Pandey-Pommier}, {Pandey}, {Pizzo}, {Polatidis}, {Rafferty}, {Rawlings}, {Reich}, {de Reijer}, {Reitsma}, {Renting}, {Riemers}, {Rol}, {Romein}, {Roosjen}, {Ruiter}, {Scaife}, {van der Schaaf}, {Scheers}, {Schellart}, {Schoenmakers}, {Schoonderbeek}, {Serylak}, {Shulevski}, {Sluman}, {Smirnov}, {Sobey}, {Spreeuw}, {Steinmetz}, {Sterks}, {Stiepel}, {Stuurwold},
  {Tagger}, {Tang}, {Tasse}, {Thomas}, {Thoudam}, {Toribio}, {van der Tol}, {Usov}, {van Veelen}, {van der Veen}, {ter Veen}, {Verbiest}, {Vermeulen}, {Vermaas}, {Vocks}, {Vogt}, {de Vos}, {van der Wal}, {van Weeren}, {Weggemans}, {Weltevrede}, {White}, {Wijnholds}, {Wilhelmsson}, {Wucknitz}, {Yatawatta}, {Zarka}, {Zensus}, \& {van Zwieten}}]{vanhaarlem2013}
{van Haarlem}, M.~P., {et~al.}, {Gunst}, A.~W., {Heald}, G. 2013, \aap, 556, A2

\bibitem[{{van Straten} \& {Bailes}(2011)}]{vanstraaten2011}
{van Straten}, W. \& {Bailes}, M. 2011, \pasa, 28, 1

\bibitem[{{van Straten} {et~al.}(2012){van Straten}, {Demorest}, \& {Oslowski}}]{straatenpsrchive}
{van Straten}, W., {Demorest}, P., \& {Oslowski}, S. 2012, Astronomical Research and Technology, 9, 237

\bibitem[{{Wang} {et~al.}(2023){Wang}, {Han}, {Xu}, {Wang}, {Yan}, {Jing}, {Su}, {Zhou}, \& {Wang}}]{whx+23}
{Wang}, P.~F., {et~al.}, {Xu}, J., {Wang}, C. 2023, Research in Astronomy and Astrophysics, 23, 104002

\bibitem[{{Weltevrede} {et~al.}(2006{\natexlab{a}}){Weltevrede}, {Edwards}, \& {Stappers}}]{waltevrede2006a}
{Weltevrede}, P., {Edwards}, R.~T., \& {Stappers}, B.~W. 2006{\natexlab{a}}, \aap, 445, 243

\bibitem[{{Weltevrede} {et~al.}(2006{\natexlab{b}}){Weltevrede}, {Stappers}, {Rankin}, \& {Wright}}]{0656rrat}
{Weltevrede}, P., {Stappers}, B.~W., {Rankin}, J.~M., \& {Wright}, G.~A.~E. 2006{\natexlab{b}}, \apjl, 645, L149

\bibitem[{{Weltevrede} {et~al.}(2006{\natexlab{c}}){Weltevrede}, {Wright}, {Stappers}, \& {Rankin}}]{walteverede2006b}
{Weltevrede}, P., {Wright}, G.~A.~E., {Stappers}, B.~W., \& {Rankin}, J.~M. 2006{\natexlab{c}}, \aap, 458, 269

\bibitem[{{Yu} {et~al.}(2020){Yu}, {Yue}, \& {Li}}]{fast_dipole}
{Yu}, J.-L., {Yue}, Y.-L., \& {Li}, J.-B. 2020, Research in Astronomy and Astrophysics, 20, 070

\bibitem[{{Zhang} {et~al.}(2024){Zhang}, {Geng}, {Wang}, {Yang}, {Kaczmarek}, {Tang}, {Johnston}, {Hobbs}, {Manchester}, {Wu}, {Jiang}, {Huang}, {Zou}, {Dai}, {Zhang}, {Li}, {Yang}, {Dai}, {Chang}, {Pan}, {Lu}, {Wei}, {Li}, {Wu}, {Qian}, {Wang}, {Wang}, {Feng}, \& {Staveley-Smith}}]{zhang_2024}
{Zhang}, S.~B., {et~al.}, {Wang}, J.~S., {Yang}, X. 2024, \apj, 972, 59

\bibitem[{Zhong {et~al.}(2024)Zhong, Spitkovsky, Mahlmann, \& Hakobyan}]{Zhong_2024}
Zhong, Y., Spitkovsky, A., Mahlmann, J.~F., \& Hakobyan, H. 2024, The Astrophysical Journal, 973, 147

\end{thebibliography}

\label{lastpage}
\end{document}